\documentclass[11pt]{article}
\usepackage[T1]{fontenc}
\usepackage{amsmath,amsfonts,amssymb}
\newcommand{\half}{{\textstyle{\frac{1}{2}}}}

\begin{document}
\newpage
\pagestyle{empty}

\vfill

\rightline{CERN-TH/2003-090}
\rightline{DSF-11/03}
\rightline{LAPTH-979/03}

\vfill

\begin{center}

{\Large \textbf{\textsf{Correlation matrix for quartet codon usage}}}

\vspace{10mm}

{\large L. Frappat$^{ad}$, A. Sciarrino$^{b}$, P. Sorba$^{ac}$}

\vspace{10mm}

\emph{$^a$ Laboratoire d'Annecy-le-Vieux de Physique Th{\'e}orique LAPTH,}

\emph{CNRS, UMR 5108, associ{\'e}e {\`a} l'Universit{\'e} de Savoie,}

\emph{BP 110, F-74941 Annecy-le-Vieux Cedex, France}

\vspace{7mm}

\emph{$^b$ Dipartimento di Scienze Fisiche, Universit{\`a} di Napoli 
``Federico II''}

\emph{and I.N.F.N., Sezione di Napoli,}

\emph{Complesso Universitario di Monte S. Angelo,} 

\emph{Via Cintia, I-80126 Naples, Italy}

\vspace{7mm}

\emph{$^c$ TH Division, CERN, CH-1211 Geneve 23, Switzerland}

\vspace{7mm}

\emph{$^d$ Member of Institut Universitaire de France}

\vspace{12mm}

\end{center}

\vspace{7mm}

\begin{abstract}
It has been argued that the sum of usage probabilities for codons, 
belonging to quartets, that have as third nucleotide $C$ or $A$, is 
independent of the biological species for vertebrates. The comparison 
between the theoretical correlation matrix derived from these sum rules and 
the experimentally computed matrix for 26 species shows a satisfactory 
agreement. The Shannon entropy, weakly depending on the biological species, 
gives further support. Suppression of codons containing the dinucleotides 
$CG$ or $AU$ is put in evidence.
\end{abstract}

\vfill
\vfill


\rightline{cond-mat/0304326}

\vspace*{3mm} 

\hrule 

\vspace*{3mm} 

\noindent 
\emph{E-mail:} \texttt{frappat@lapp.in2p3.fr}, 
\texttt{sciarrino@na.infn.it}, \texttt{sorba@lapp.in2p3.fr}, 
\texttt{sorba@cern.ch} (on leave of absence from LAPTH)

\newpage
\pagestyle{plain}
\setcounter{page}{1}
\baselineskip=16pt

\section{Introduction}

The genetic information in DNA is stored in sequences built up from four 
bases (nucleotides) $C$, $T$, $G$, $A$ (in mRNA, which plays a key role in 
the construction of proteins, the base $T$ is replaced by $U$). The 
proteins are made up from 20 different amino-acids. The ``quantum'' of 
genetic information is constituted by an ordered triplet of nucleotides 
(codon). There are therefore 64 possible codons, which encode 20 
amino-acids, plus the three signals of the termination of the biosynthesis 
process. It follows that the genetic code, i.e. the correspondence between 
codons and amino-acids, is degenerate, and almost all the amino-acids are 
encoded by multiple codons (synonymous codons). Degeneracy is mainly found 
in the third position of the codon. For a long time it was accepted that: 
1) there was no preference for the use of any particular codon, except for 
the enhancement due to the abundance of a particular basis (nucleotide); 2) 
the codon usage frequency was strongly dependent on the considered 
biological species, and no regular pattern could therefore be extracted.

\medskip

The currently available data show that some codons are used much more 
frequently than others to encode a particular amino-acid. Let us, indeed, 
define the usage probability for the codon $XZN$ ($X, Z, N \in \{A,C, G, 
U\}$) in a quartet as
\begin{equation}
\label{eq:1}
P(XZN) = \lim_{n_{tot} \to \infty} \;\;\; \frac{n_{XZN}}{n_{XZ}}
\end{equation}
where $n_{XZN}$ is the number of times the codon $XZN$ has been used in the 
biosynthesis process of the corresponding amino-acid and $n_{XZ}$ is the 
total number of codons used to synthetize this amino-acid. It follows that 
our analysis and predictions hold for biological species with sufficiently 
large statistics of codons. The above definition holds also for the usage 
probability for doublets, in which case we have to consider two 
possibilities: $ N \equiv Y (= C,U)$ and $ N \equiv R (= G,A)$. Let us 
point out that we consider a sextet as the sum of a quartet and a doublet, 
and a triplet as the sum of a doublet and a singlet. Therefore, in the 
following, we consider eight quartets, corresponding to the five quartets 
plus the quartet subparts of the three sextets of the eukaryotic code. In 
the following the probabilities for each quartet or each doublet are 
normalized to 1. It is currently believed that a non-uniform usage of 
synonymous codons is a widespread phenomenon and it is experimentally 
observed that the pattern of codon usage varies between species and even 
between tissues within a species; see refs. \cite{dumou99,kana2001}, which 
contain a large number of references to the original works on the subject. 
The main reasons for the codon usage biases are believed to be: the 
mutational biases, the translation efficiency, the natural selection and 
the abundance of specific anticodons in the tRNA. However, the aim of this 
paper is not to compare the different proposed explanations, but to put in 
evidence a general pattern of the bias.

\medskip

To our knowledge, systematic studies of codon usage for eukaryotes are 
rather fragmentary, while the case of bacteriae has been widely studied. 
Most of the analyses of the codon usage frequencies have addressed an 
analysis of the relative abundance of a specified codon in different genes 
of the same biological species or in the comparison of the relative 
abundance in the same gene for different biological species. Little 
attention has been paid to analyze the codon usage frequency summed over 
the whole available sequences to infer global correlations between 
different biological species. Indeed, in \cite{KFL}, analysing a large 
sample of species, a correlation between the GC content and the codon usage 
has been pointed out and explained on the basis of a mutational model at 
the nucleotide level. In \cite{FSS2002}, starting from the so-called 
crystal basis model for the genetic code proposed a few years ago by the 
present authors in \cite{FSS}, it has been derived that the sum of the 
codon usage probabilities for codons belonging to quartets, in the 
generalized meaning specified above, with third nucleotide $C$ and $A$ (or 
$G$ and $U$), should be a constant (sum rule). In the present paper we 
pursue the analysis and the developments of these sum rules, in particular 
investigating the consequences on the correlation matrix. With the twofold 
aim of focussing on the implications of the sum rules and of making the 
paper more easily legible to a wide range of readers, which may be not 
acquainted with the mathematical background of our model, no reference will 
be made either to the model itself or to the mathematical reasoning that 
led to the derivation of the sum rules. The interested reader can find 
details in the quoted papers and references therein. We simply state the 
results, which will be the input for the further analysis presented below.

\textbf{Sum rules:} for the codons belonging to quartets and for biological 
species belonging to vertebrates,
\begin{eqnarray}
\label{eq:11}
P(XZC) + P(XZA) = \mbox{Const.} \qquad (XZ = NC, CU, GU, CG, GG)
\end{eqnarray}
holds. Here Const. means a value, depending on $XZ$, but independent on 
the biological species. In \cite{FSS2002} a table with the values of these 
quantities can be found. From (\ref{eq:11}), one gets immediately
\begin{eqnarray}
P(XZC) & = & \rho(XZC) \, + \, H_{bs}(XZC) \nonumber \\
P(XZA) & = & \rho(XZA) \, - \, H_{bs}(XZC) \nonumber \\
P(XZG) & = & \rho(XZG) \, + \, H_{bs}(XZG) \nonumber \\
P(XZU) & = & \rho(XZU) \, - \, H_{bs}(XZG) \;,
\label{eq:prob}
\end{eqnarray}
where $\rho$ are functions independent on the biological species while the 
functions $H_{bs}$ depend on the biological species.

\medskip

In \cite{FSS2002} the r.h.s. of eq. (\ref{eq:11}) has been fitted to the 
experimental data. Therefore it is natural to wonder how the experimentally 
fitted values depend on the data. In table \ref{tablebs} we have computed, 
for two biological species, namely \textit{Homo sapiens} and \textit{Danio 
rerio}, the probabilities for the codons in quartets with the last 
nucleotide $C$ or $A$ and their sum, using different releases of available 
data from GenBank in the last four years, in order to investigate the 
dependence on the statistics. One can remark that the experimentally 
computed probabilities change, but that in about $80 \%$ of cases an 
increase in $P(XZA)$ corresponds to a decrease in $P(XZC)$ and vice versa. 
This property is an indication of the existence of an anticorrelation 
between $P(XZC)$ and $P(XZA)$. An analogous analysis for other biological 
species belonging to the sample listed in table \ref{tabledata1} shows the 
same behaviour. Let us remark that from this kind of analysis, one expects 
that for statistics of codons lower than roughly 100\,000, the statistical 
fluctuations become non-negligible.

\section{Correlation matrix}

Let us introduce now the correlation matrix $\Gamma$ between two discrete 
random variables $X$ and $Y$. For a sample of size $n$, it is defined by
\begin{equation}
\label{eq:correl}
\Gamma(X,Y) = \frac{1} {\sigma_{X} \, \sigma_{Y}} \; \sum_{k=1}^{n} 
\frac{1}{n} \; \Big( X_{k} - \langle X \rangle \Big) \Big( Y_{k} - \langle 
Y \rangle \Big)
\end{equation}
where $\sigma_{X}$ is the standard deviation
\begin{equation}
\label{eq:stdev}
\sigma_{X} = \sqrt{ \frac{1}{n(n-1)} \left( n \sum_{k=1}^{n} \, X_{k}^{2} - 
\Big(\sum_{k=1}^{n} \, X_{k}\Big)^{2} \right)}
\end{equation}
and $\langle X \rangle$ is the mean value
\begin{equation}
\label{eq:meanval}
\langle X \rangle = \frac{1}{n} \; \sum_{k=1}^{n} X_{k} \;.
\end{equation}
In our context, $X_{k}$ (resp. $Y_{k}$) are the usage probability functions 
$P(XZN)$ (resp. $P(X'Z'N')$) for the $k$-th biological species for the same 
or different amino-acids. The sample we consider is given in table 
\ref{tabledata1}.

We can then construct a $32 \times 32$ correlation matrix $\Gamma$ whose 
entries are the correlation coefficients $\Gamma(P(XZN),$ $P(X'Z'N'))$, 
where $XZ,X'Z' = CC,UC,GC,AC,CU,GU,CG,GG$. From the expressions of the 
probabilities (\ref{eq:1}) and the sum rules (\ref{eq:11}) we derive for 
the same amino-acid the following symmetric submatrix:
\[
\begin{array}{|c|c|c|c|c|}
\hline
& P(XZU) & P(XZC) & P(XZA) & P(XZG) \\
\hline
P(XZU) &  1 & -x &  x & -1 \\
P(XZC) & -x &  1 & -1 &  x \\
P(XZA) &  x & -1 &  1 & -x \\
P(XZG) & -1 &  x & -x &  1 \\
\hline
\end{array}
\]
where $x$ is a theoretically undetermined quantity to be fitted 
experimentally for any amino-acid. This kind of pattern can be easily 
compared with the experimental data for vertebrates (see the list of 
biological species and the corresponding number of codons used) from 
GenBank (release 131.0 of August 2002). One gets for the eight quartets

\[
\begin{array}{|l||c|c|c|c|c|c|}
\hline
& \Gamma_{UC} & \Gamma_{UA} & \Gamma_{UG} & \Gamma_{CA} & \Gamma_{CG} & 
\Gamma_{AG} \\
\hline
\mbox{Pro} & -0.75 & \phantom{-}0.71 & -0.68 & -0.88 & \phantom{-}0.22 & 
-0.53 \\
\mbox{Thr} & -0.89 & \phantom{-}0.88 & -0.62 & -0.89 & \phantom{-}0.27 & 
-0.62 \\
\mbox{Ala} & -0.86 & \phantom{-}0.70 & -0.46 & -0.80 & \phantom{-}0.17 & 
-0.61 \\
\mbox{Ser} & -0.75 & \phantom{-}0.44 & -0.67 & -0.84 & \phantom{-}0.30 & 
-0.40 \\
\mbox{Val} & -0.55 & \phantom{-}0.51 & -0.89 & -0.67 & \phantom{-}0.32 & 
-0.66 \\
\mbox{Leu} & -0.83 & \phantom{-}0.43 & -0.82 & -0.47 & \phantom{-}0.57 & 
-0.78 \\
\mbox{Arg} & -0.36 & \phantom{-}0.34 & -0.92 & -0.88 & \phantom{-}0.03 & 
-0.16 \\
\mbox{Gly} & -0.78 & \phantom{-}0.78 & -0.91 & -0.92 & \phantom{-}0.63 & 
-0.80 \\
\hline
\end{array}
\]
where $\Gamma_{NN'}$ denotes for short $\Gamma(P(XZN),P(XZN'))$, for some 
amino-acid related to a quartet of codons $XZN$.

{From} the above table, we remark that the values of some antidiagonal 
elements $\Gamma_{NN'}$ are very close to the theoretical values $-1$, 
namely $\Gamma_{UG}$ and $\Gamma_{CA}$ for Arg and Gly, $\Gamma_{CA}$ for 
Pro, Thr, Ala, Ser and $\Gamma_{UG}$ for Val and Leu \footnote{In the 
crystal basis model \cite{FSS}, the mathematical assignment of the codons 
encoding for the eight quartets exhibit the same pattern of grouping of 
codons.}. We also note that the entries $\Gamma_{UN}$ for all eight 
amino-acids are in good agreement with the theoretical values, while the 
entries $\Gamma_{CG}$, $\Gamma_{AG}$ show a large numerical discrepancy. 
These facts mean that the small violations of the sum rules are averaged, 
in the sum defining the correlation coefficients, almost to zero for 
$\Gamma_{UN}$, but not for the other ones, as can be inferred from the 
analysis of the values of the probabilities $P(XZN)$ given in tables 
\ref{tablePC} to \ref{tablePG}. It is evident that the degree of 
correlation is not the same for the eight amino-acids, Thr and Leu being 
the most correlated ones, and Arg the least one.

It may be useful to compute the average value over the eight amino-acids of 
the entries of the correlation matrix
\begin{align}
& \langle\Gamma_{CA}\rangle = -0.79 \;, && \langle\Gamma_{CG}\rangle = 0.32 
\;, && \langle\Gamma_{CU}\rangle = -0.72 \;, \nonumber \\
& \langle\Gamma_{GU}\rangle = -0.75 \;, && \langle\Gamma_{AU}\rangle = 0.60 
\;, && \langle\Gamma_{AG}\rangle = -0.57 \;.
\end{align}
We see that the average values differ within $20 \%$ from the expected 
ones, except for $\Gamma_{CG}$, which is almost half of the expected value. 
We shall comment later on this point.

\medskip

For different amino-acids we derive from (\ref{eq:prob}) that for the eight 
amino-acids related to quartets, the theoretical correlation pattern, for 
the different probabilities take the following form
\[
\begin{array}{|c|c|c|c|c|}
\hline
& P(XZU) & P(XZC) & P(XZA) & P(XZG) \\
\hline
P(X'Z'U) &  y & -z &  z & -y \\
P(X'Z'C) & -t &  x & -x &  t \\
P(X'Z'A) &  t & -x &  x & -t \\
P(X'Z'G) & -y &  z & -z &  y \\
\hline
\end{array}
\]
This kind of pattern can also be easily compared with the experimental 
data. One obtains, for example, for the correlations between the 
probabilities of codon usage of Valine and Threonine:
\[
\begin{array}{|c|c|c|c|c|}
\hline
 & P(GUU) & P(GUC) & P(GUA) & P(GUG) \\
\hline
P(ACU) & \phantom{-}0.76 & -0.64 & \phantom{-}0.75 & -0.76 \\
P(ACC) & -0.83 & \phantom{-}0.64 & -0.72 & \phantom{-}0.81 \\
P(ACA) & \phantom{-}0.72 & -0.80 & \phantom{-}0.83 & -0.67 \\
P(ACG) & -0.23 & \phantom{-}0.57 & -0.57 & \phantom{-}0.22 \\
\hline
\end{array}
\]
for Glycine and Alanine:
\[
\begin{array}{|c|c|c|c|c|}
\hline
 & P(GGU) & P(GGC) & P(GGA) & P(GGG) \\
\hline
P(GCU) & \phantom{-}0.83 & -0.80 & \phantom{-}0.71 & -0.67 \\
P(GCC) & -0.65 & \phantom{-}0.86 & -0.80 & \phantom{-}0.54 \\
P(GCA) & \phantom{-}0.56 & -0.79 & \phantom{-}0.69 & -0.40 \\
P(GCG) & -0.48 & \phantom{-}0.35 & -0.22 & \phantom{-}0.31 \\
\hline
\end{array}
\]

We can extend the correlation matrix by taking into account the 
correlations between quartets and doublets, getting a $58 \!\times\! 58$ 
matrix. We obtain, for the correlation entries between quartets and 
doublets:
\[
\begin{array}{|c|c|c|c|c|}
\hline
& P(XZU) & P(XZC) & P(XZA) & P(XZG) \\
\hline
P(X'Z'U) &  x & -y &  y & -x \\
P(X'Z'C) & -x &  y & -y &  x \\
\hline
P(X'Z'A) &  z & -t &  t & -z \\
P(X'Z'G) & -z &  t & -t &  z \\
\hline
\end{array}
\]
Note that the structure of the doublets implies that the second and fourth 
lines are direct consequences of the first and third lines, respectively. 
As an example, we obtain, for the correlations between Pro versus Asp and 
Glu:
\[
\begin{array}{|c|c|c|c|c|}
\hline
 & P(CCU) & P(CCC) & P(CCA) & P(CCG) \\
\hline
P(GAU) & \phantom{-}0.68 & -0.76 & \phantom{-}0.88 & -0.55 \\
\hline
P(GAA) & \phantom{-}0.54 & -0.58 & \phantom{-}0.75 & -0.54 \\
\hline
\end{array}
\]
Also in this case, the largest discrepancy appears in the matrix entry 
$\Gamma_{CG}$.

\medskip

An analysis of the above tables shows that there are numerical violations 
with respect to the theoretical correlation matrix, but that the signs, 
with few exceptions corresponding to values close to zero, are preserved. 
Let us remark that for each pair of quartets (or quartet and doublet), 
provided that the codons are put in the order $U, C, A, G$ and assuming 
that the number $x,y,z,t$ are all of the same sign, one obtains a 
correlation matrix whose entries show alternating signs along each row and 
each column. An inspection of the correlation matrix shows that the 406 
independent parameters that appear in our model are all positive. This 
intriguing property, which is likely to reflect some ``dynamical'' 
behaviour, requires further analysis in order to be well understood and 
also to evaluate to what extent it depends on the choice of the sample. A 
comparison with the experimentally computed $58 \!\times\! 58$ correlation 
matrix shows that this property is verified with a very good approximation. 
Indeed it is violated only for a few per cent of the entries: in these 
cases the correlation coefficients with wrong sign are very small, i.e. 
they concern poorly correlated codon usage probabilities, which are 
expected to be much more sensitive to fluctuations.

\section{Shannon entropy}

The order of magnitude of the biological species independent term 
$\rho(XZN)$ in (\ref{eq:prob}) is around the mean value denoted $\langle 
P_{N} \rangle$ in tables \ref{tablePC} to \ref{tablePG}, and the biological 
species dependent term in (\ref{eq:prob}) is around $\sigma(P_{N})$ 
($N=C,A,U,G$). Note that the ratio $\sigma(P_{N})/\langle P_{N} \rangle$ is 
around 10\% in each case. A rough analysis of the distribution of the 
probabilities $P_{N}$ seems to indicate a standard distribution, although 
the small size of the sample does not allow us to draw a reliable 
conclusion. \\
In order to check the effectiveness of the sum rules in the light of the 
above estimations, we compute the Shannon entropy related to the eight 
amino-acids, all considered as encoded by quartets:
\begin{equation}
	S_{XZ} = -\sum_{N=C,A,U,G} P(XZN) \, \ln_{2} P(XZN)
	\label{eq:shan}
\end{equation}
where $P(XZN)$ is given by eq. (\ref{eq:prob}). Since 
${H_{bs}}_{N}/\rho_{N} \ll 1$, we can approximate $\ln (1 \pm 
{H_{bs}}_{N}/\rho_{N})$ by $\pm {H_{bs}}_{N}/\rho_{N} - \half 
{H_{bs}^2}_{N}/\rho_{N}^2$.
So we get for the Shannon entropy for a given amino-acid related 
to a dinucleotide $XZ$:
\begin{eqnarray}
S_{XZ} &\simeq& -\sum_{N=C,A,U,G} \rho_{N} \, \ln_{2} \rho_{N} - 
{H_{bs}}_{C} \, \ln_{2} \frac{\rho_{C}}{\rho_{A}} - {H_{bs}}_{G} \, \ln_{2} 
\frac{\rho_{G}}{\rho_{U}} \nonumber \\
&-& \frac{1}{2\ln 2} \; {H_{bs}^2}_{C} 
\left(\frac{1}{\rho_{C}}+\frac{1}{\rho_{A}}\right) - \frac{1}{2\ln 2} \; 
{H_{bs}^2}_{G} \left(\frac{1}{\rho_{G}}+\frac{1}{\rho_{U}}\right) \;,
\label{eq:entropy}
\end{eqnarray}
where for simplicity $\rho_{N}$, ${H_{bs}}_{N}$ and $P_{N}$ stand for 
$\rho(XZN)$, $H_{bs}(XZN)$ and $P(XZN)$ respectively, the dinucleotide $XZ$ 
being fixed. In this equation, the first sum corresponds to the biological 
species independent part of the Shannon entropy, while the second and third 
terms are the first- and second-order biological species dependent 
corrections. From the above formula, we expect the entropy to be 
independent of the biological species within a few per cent. We report in 
table \ref{shannon} the Shannon entropy for the eight amino-acids related 
to quartets for the sample of biological species of table \ref{tabledata1}. 
The standard deviation is in this case reduced with respect to the standard 
deviations of the probability distributions $P(XZN)$, see tables 
\ref{tablePC} to \ref{tablePG}, as can be easily checked quantitatively 
from the last terms of (\ref{eq:entropy}). Let us remark that there is no 
clear dependence of the Shannon entropies $S_{XZ}$ on the total exonic $GC$ 
content of the biological species.

\medskip

Now we can compute the contribution of each quartet to the Shannon entropy 
of the total exonic region, defined as:
\begin{equation}
S_{tot} = -\sum_{i} p_{i} \, \ln_{2} p_{i} \;,
\label{eq:shannon}
\end{equation}
where the sum runs over all the coding codons and $p_{i}$ is the usage 
probability of the codon $i$, i.e.
\begin{equation}
\label{eq:pi}
p_{i} = \frac{n_{i}}{N_{tot}}
\end{equation}
where $n_{i}$ is the number of times the codons $i$ has been used and 
$N_{tot}$ is the total number of coding codons. \\
The contribution to the Shannon entropy (\ref{eq:shannon}) of the 
amino-acid encoded by the codons $XZN$ ($N=C,A,U,G$) is
\begin{equation}
\label{eq:shanpart}
S_{tot}^{XZ} = - \sum_{N=C,A,U,G} p_{N} \, \ln_{2} p_{N} \;.
\end{equation}
As $\displaystyle p_{N} = P(XZN) \, \frac{n_{XZ}}{N_{tot}}$, we get
\begin{equation}
\label{eq:shanpart2}
S_{tot}^{XZ} = - P_{XZ} \, \ln_{2} P_{XZ} + P_{XZ} \, S_{XZ} \;,
\end{equation}
where $P_{XZ}$ is the probability of the amino-acid encoded by $XZN$, i.e. 
$\displaystyle P_{XZ} = \frac{n_{XZ}}{N_{tot}}$, and $S_{XZ}$ is given by 
eq. (\ref{eq:shan}).

Therefore, from the remarked biological species independence of $S_{XZ}$, 
we derive that the contribution of an amino-acid (encoded by a quartet) to 
the total Shannon entropy depends only on its frequency. This is a non 
trivial result. Indeed, the probability of an amino-acid depends of course 
on the probabilities of the encoding codons, but a priori one might obtain 
the same amino-acid probability with different distributions of the 
encoding codons, therefore with different contributions to the total 
Shannon entropy. On the contrary, eq. (\ref{eq:shanpart2}) tells us that 
the contribution to the entropy is, within a few per cent, equal for all 
vertebrates for a fixed frequency of the amino-acid.

\section{Conclusion}

The above analysis provides a strong support of the validity for the sum 
rules equation (\ref{eq:11}). The emerging pattern, for vertebrates, is the 
existence of a species independent codon usage bias. From tables 
\ref{tablePC} to \ref{tablePG}, it is clear that the third nucleotides in 
the quartets are in decreasing usage order: $C > U > A > G$ for Pro, Ala 
and Ser, $C > U > G > A$ for Thr, $G > C > U > A$ for Val and Leu, $C 
\gtrsim G > A \gtrsim U$ for Arg and $C > A > G > U$ for Gly. Some 
biological species show different patterns (e.g. \textit{Danio rerio} and 
\textit{Macaca fascicularis}). The deviations from the values given by the 
sum rules are fluctuations that shift the numerical values of the 
correlation matrix entries from the expected ones, but are not strong 
enough to change the sign. The existence of sum rules indicates 
correlations in the whole set of the coding sequences of DNA, while in the 
literature it is stated that correlations essentially appear in the 
non-coding part of DNA. It should be quoted that Zeeberg \cite{Zee}, from 
an analysis of over 7000 genes for \textit{Homo sapiens}, has derived 
parameters for linear regression for codon nucleotide composition versus 
the per cent of $GC$ content in genomic regions. From his analysis, one 
expects a correlation of the same order for codons having $C$ or $U$, and 
$C$ or $A$ in third position. Our result shows a stronger correlation in 
the latter case.

\medskip

As noted above, the values of the entry $\Gamma_{CG}$ are smaller than 
those expected. This feature may be due to the known suppression of the 
codons containing a dinucleotide CG, see ref. \cite{kana2001}, and for a 
thermodynamical explanation of this phenomenon, see ref. \cite{klump91}. 
Note that this suppression can be inferred by looking at the first four 
columns of table \ref{tablePG}. An inspection of the columns 5 and 6 of 
table \ref{tablePA} shows that there is also a suppression in the usage of 
codons containing the dinucleotide $UA$. It is worthwhile to further 
investigate this suppression in the other codons. Indeed an analysis of the 
codon usage probabilities of the codons $AUA$ encoding for Ile and $UUA$ 
encoding for Leu confirms this suppression. It is also interesting to 
observe that the codons $UAR$, being nonsense codons, are rarely used.

\medskip

We have also remarked that the Shannon entropies related to the eight 
amino-acids, encoded by quartets in the present scheme, are biological 
species independent within a few per cent. Moreover, the contribution of 
these amino-acids to the total Shannon entropy (\ref{eq:shannon}) depends 
only on their frequencies.

\medskip

It would be interesting to extend this type of analysis to a suitably 
chosen sample of plants or of invertebrates, while bacteriae seem to behave 
in a peculiar different way.

\bigskip

\textbf{Acknowledgements:} A.S. is indebted to the Universit{\'e} de Savoie 
for financial support and LAPTH for its kind hospitality.

\begin{table}[htbp]
\centering
\caption{Data for the biological species (release 131.0 of GenBank)
\label{tabledata1}}
\begin{tabular}{|r|l|r|r|c|}
\hline
& Species & $\#$ of seqs. & $\#$ of codons & exonic $GC$ content \\
\hline
1 & Homo sapiens & 55194 \; & 24298072 \; & 52.45 \\
2 & Mus musculus & 25249 \; & 11455875 \; & 52.38 \\
3 & Rattus norvegicus & 7260 \; & 3637311 \; & 52.75 \\
4 & Xenopus laevis & 3057 \; & 1402544 \; & 47.25 \\
5 & Gallus gallus & 2352 \; & 1152438 \; & 52.05 \\
6 & Danio rerio & 2535 \; & 1124891 \; & 50.91 \\
7 & Bos taurus & 2012 \; & 884632 \; & 52.39 \\
8 & Sus scrofa & 1256 \; & 503449 \; & 53.44 \\
9 & Macaca fascicularis & 1547 \; & 466424 \; & 49.60 \\
10 & Oryctolagus cuniculus & 914 \; & 462701 \; & 54.40 \\
11 & Canis familiaris & 576 \; & 270095 \; & 52.19 \\
12 & Takifugu rubripes & 373 \; & 212447 \; & 54.24 \\
13 & Oncorhynchus mykiss & 454 \; & 163360 \; & 53.53 \\
14 & Cavia porcellus & 360 \; & 144107 \; & 51.73 \\
15 & Ovis aries & 444 \; & 143625 \; & 53.46 \\
16 & Oryzias latipes & 265 \; & 119801 \; & 51.59 \\
17 & Cricetulus griseus & 247 \; & 119508 \; & 51.37 \\
18 & Rattus sp. & 277 \; & 113465 \; & 52.50 \\
19 & Macaca mulatta & 364 \; & 109818 \; & 51.43 \\
20 & Mesocricetus auratus & 246 \; & 108131 \; & 52.41 \\
21 & Pan troglodytes & 325 \; & 104247 \; & 57.49 \\
22 & Cyprinus carpio & 250 \; & 103536 \; & 49.60 \\
23 & Mus sp. & 236 \; & 92973 \; & 52.87 \\
24 & Felis catus & 222 \; & 81469 \; & 52.15 \\
25 & Equus caballus & 221 \; & 77391 \; & 52.97 \\
26 & Rattus rattus & 179 \; & 71784 \; & 53.11 \\
\hline
\end{tabular}
\end{table}

\begin{table}[htbp]
\centering
\caption{ Number of codons for \textit{Homo sapiens}, \textit{Danio rerio} 
and \textit{Macaca fascicularis} in different releases from GenBank used to 
compute probabilities reported in table \ref{tablebs}.}
\label{tablerel}
\footnotesize 
\medskip
\begin{tabular}{|l||c|c||c|c||c|c|}
\hline
Homo sapiens & 1 & 6\,130\,940 & 2 & 8\,707\,603 & 3 & 11\,310\,862 \\
\hline
Homo sapiens & 4 & 19\,894\,411 & 5 & 24\,298\,072 & &\\
\hline
\hline
Danio rerio & 1 & 99\,766 & 2 & 213\,258 & 3 & 312\,789 \\
\hline
Danio rerio & 4 & 696\,043 & 5 & 1\,124\,891 & & \\
\hline
\end{tabular}
\end{table}

\clearpage

\begin{table}[htbp]
\centering
\caption{Probabilities for $P(XZC)$, $P(XZA)$ and $P(XZC) + P(XZA)$ 
corresponding to the eight amino-acids related to quartets, for 
\textit{Homo sapiens}, \textit{Danio rerio} and \textit{Macaca 
fascicularis}, computed using data from different releases from 1998 to 
2002, corresponding to a total number of codons reported in table 
\ref{tablerel}.}
\label{tablebs}
\footnotesize \medskip
\begin{tabular}{|c||c|c|c|c|c|c|c|c|}
\hline
\multicolumn{9}{|c|}{\textit{Homo sapiens}} \\
\hline
release & $P_{C}(P)$ & $P_{C}(T)$ & $P_{C}(A)$ & $P_{C}(S)$ & $P_{C}(V) $ & 
$P_{C}(L)$ & $P_{C}(R)$ & $P_{C}(G)$ \\
\hline
1 & 0.3336 & 0.3785 & 0.4109 & 0.3743 & 0.2462 & 0.2497 & 0.3332 & 0.3480 
\\
2 & 0.3318 & 0.3699 & 0.4072 & 0.3671 & 0.2400 & 0.2462 & 0.3287 & 0.3453 
\\
3 & 0.3284 & 0.3640 & 0.4040 & 0.3618 & 0.2387 & 0.2450 & 0.3243 & 0.3430 
\\
4 & 0.3280 & 0.3620 & 0.4037 & 0.3612 & 0.2380 & 0.2446 & 0.3228 & 0.3435 
\\
5 & 0.3300 & 0.3619 & 0.4026 & 0.3609 & 0.2386 & 0.2461 & 0.3221 & 0.3415 
\\
\hline
\hline
release & $P_{A}(P)$ & $P_{A}(T)$ & $P_{A}(A)$ & $P_{A}(S)$ & $P_{A}(V) $ & 
$P_{A}(L)$ & $P_{A}(R)$ & $P_{A}(G)$ \\
\hline
1 & 0.2705 & 0.2691 & 0.2189 & 0.2334 & 0.1063 & 0.0835 & 0.1811 & 0.2454 
\\
2 & 0.2705 & 0.2737 & 0.2215 & 0.2369 & 0.1107 & 0.0863 & 0.1850 & 0.2453 
\\
3 & 0.2734 & 0.2778 & 0.2251 & 0.2417 & 0.1133 & 0.0878 & 0.1883 & 0.2478 
\\
4 & 0.2737 & 0.2788 & 0.2256 & 0.2427 & 0.1138 & 0.0875 & 0.1875 & 0.2461 
\\
5 & 0.2734 & 0.2817 & 0.2280 & 0.2462 & 0.1175 & 0.0970 & 0.1895 & 0.2479 
\\
\hline
\hline
release & $P_{C+A}(P)$ & $P_{C+A}(T)$ & $P_{C+A}(A)$ & $P_{C+A}(S)$ & 
$P_{C+A}(V) $ & $P_{C+A}(L)$ & $P_{C+A}(R)$ & $P_{C+A}(G)$ \\
\hline
1 & 0.6041 & 0.6476 & 0.6298 & 0.6077 & 0.3525 & 0.3332 & 0.5143 & 0.5934 
\\
2 & 0.6023 & 0.6436 & 0.6287 & 0.6040 & 0.3507 & 0.3325 & 0.5137 & 0.5906 
\\
3 & 0.6018 & 0.6418 & 0.6291 & 0.6035 & 0.3520 & 0.3328 & 0.5126 & 0.5908 
\\
4 & 0.6016 & 0.6408 & 0.6293 & 0.6039 & 0.3518 & 0.3321 & 0.5103 & 0.5896 
\\
5 & 0.6034 & 0.6436 & 0.6306 & 0.6071 & 0.3561 & 0.3431 & 0.5116 & 0.5894 
\\
\hline
\hline
\multicolumn{9}{|c|}{\textit{Danio rerio}} \\
\hline
release & $P_{C}(P)$ & $P_{C}(T)$ & $P_{C}(A)$ & $P_{C}(S)$ & $P_{C}(V) $ & 
$P_{C}(L)$ & $P_{C}(R)$ & $P_{C}(G)$ \\
\hline
1 & 0.2717 & 0.3445 & 0.3202 & 0.3329 & 0.2457 & 0.2420 & 0.3432 & 0.3010 
\\
2 & 0.2708 & 0.3392 & 0.3244 & 0.3407 & 0.2495 & 0.2510 & 0.3407 & 0.3056 
\\
3 & 0.2641 & 0.3344 & 0.3208 & 0.3331 & 0.2468 & 0.2434 & 0.3379 & 0.2981 
\\
4 & 0.2601 & 0.3212 & 0.3168 & 0.3198 & 0.2387 & 0.2406 & 0.3326 & 0.2872 
\\
5 & 0.2542 & 0.3117 & 0.3113 & 0.3133 & 0.2376 & 0.2373 & 0.3271 & 0.2845 
\\
\hline
\hline
release & $P_{A}(P)$ & $P_{A}(T)$ & $P_{A}(A)$ & $P_{A}(S)$ & $P_{A}(V) $ & 
$P_{A}(L)$ & $P_{A}(R)$ & $P_{A}(G)$ \\
\hline
1 & 0.2666 & 0.2629 & 0.2347 & 0.2383 & 0.1010 & 0.0891 & 0.2142 & 0.3239 
\\
2 & 0.2629 & 0.2696 & 0.2327 & 0.2363 & 0.0990 & 0.0878 & 0.2140 & 0.3219 
\\
3 & 0.2718 & 0.2738 & 0.2373 & 0.2379 & 0.0998 & 0.0833 & 0.2141 & 0.3285 
\\
4 & 0.2749 & 0.2854 & 0.2417 & 0.2437 & 0.1049 & 0.0836 & 0.2195 & 0.3379 
\\
5 & 0.2816 & 0.2942 & 0.2461 & 0.2477 & 0.1035 & 0.0832 & 0.2252 & 0.3413 
\\
\hline
\hline
release & $P_{C+A}(P)$ & $P_{C+A}(T)$ & $P_{C+A}(A)$ & $P_{C+A}(S)$ & 
$P_{C+A}(V) $ & $P_{C+A}(L)$ & $P_{C+A}(R)$ & $P_{C+A}(G)$ \\
\hline
1 & 0.5383 & 0.6075 & 0.5548 & 0.5712 & 0.3466 & 0.3311 & 0.5574 & 0.6249 
\\
2 & 0.5337 & 0.6088 & 0.5571 & 0.5770 & 0.3485 & 0.3388 & 0.5547 & 0.6275 
\\
3 & 0.5359 & 0.6081 & 0.5582 & 0.5709 & 0.3466 & 0.3267 & 0.5520 & 0.6266 
\\
4 & 0.5350 & 0.6066 & 0.5584 & 0.5634 & 0.3436 & 0.3242 & 0.5522 & 0.6250 
\\
5 & 0.5358 & 0.6059 & 0.5575 & 0.5610 & 0.3410 & 0.3205 & 0.5523 & 0.6257 
\\
\hline
\end{tabular}
\end{table}

\clearpage

\begin{table}[htbp]
\centering
\caption{Probabilities $P(XZC)$ for the eight amino-acids related to 
quartets (release 131.0 of GenBank database). First column 
refers to the biological species of table \ref{tabledata1}.}
\label{tablePC}
\footnotesize \medskip
\begin{tabular}{|c||c|c|c|c|c|c|c|c|}
\hline
& $P_{C}(P)$ & $P_{C}(T)$ & $P_{C}(A)$ & $P_{C}(S)$ & $P_{C}(V)$ & 
$P_{C}(L)$ & $P_{C}(R)$ & $P_{C}(G)$ \\
\hline
1 & 0.330 & 0.362 & 0.403 & 0.361 & 0.239 & 0.246 & 0.322 & 0.342 \\
2 & 0.306 & 0.354 & 0.383 & 0.363 & 0.250 & 0.248 & 0.306 & 0.334 \\
3 & 0.319 & 0.372 & 0.401 & 0.380 & 0.261 & 0.255 & 0.313 & 0.340 \\
4 & 0.235 & 0.273 & 0.278 & 0.301 & 0.199 & 0.197 & 0.262 & 0.236 \\
5 & 0.331 & 0.333 & 0.347 & 0.359 & 0.231 & 0.234 & 0.364 & 0.321 \\
6 & 0.254 & 0.312 & 0.311 & 0.313 & 0.238 & 0.237 & 0.327 & 0.284 \\
7 & 0.350 & 0.383 & 0.427 & 0.377 & 0.257 & 0.255 & 0.329 & 0.349 \\
8 & 0.356 & 0.414 & 0.448 & 0.419 & 0.266 & 0.255 & 0.354 & 0.366 \\
9 & 0.292 & 0.315 & 0.363 & 0.328 & 0.229 & 0.242 & 0.289 & 0.300 \\
10 & 0.385 & 0.416 & 0.474 & 0.447 & 0.277 & 0.269 & 0.393 & 0.392 \\
11 & 0.350 & 0.379 & 0.424 & 0.376 & 0.264 & 0.259 & 0.341 & 0.334 \\
12 & 0.339 & 0.401 & 0.409 & 0.418 & 0.305 & 0.266 & 0.363 & 0.332 \\
13 & 0.384 & 0.430 & 0.416 & 0.404 & 0.283 & 0.251 & 0.381 & 0.298 \\
14 & 0.335 & 0.380 & 0.413 & 0.392 & 0.245 & 0.251 & 0.364 & 0.345 \\
15 & 0.369 & 0.431 & 0.452 & 0.401 & 0.266 & 0.277 & 0.371 & 0.361 \\
16 & 0.293 & 0.372 & 0.351 & 0.372 & 0.269 & 0.234 & 0.367 & 0.300 \\
17 & 0.320 & 0.379 & 0.377 & 0.361 & 0.243 & 0.235 & 0.296 & 0.342 \\
18 & 0.316 & 0.382 & 0.401 & 0.373 & 0.274 & 0.256 & 0.309 & 0.344 \\
19 & 0.318 & 0.384 & 0.403 & 0.389 & 0.262 & 0.268 & 0.334 & 0.310 \\
20 & 0.332 & 0.378 & 0.404 & 0.376 & 0.257 & 0.237 & 0.308 & 0.323 \\
21 & 0.415 & 0.496 & 0.388 & 0.419 & 0.242 & 0.273 & 0.418 & 0.367 \\
22 & 0.259 & 0.312 & 0.315 & 0.309 & 0.247 & 0.233 & 0.298 & 0.273 \\
23 & 0.296 & 0.358 & 0.395 & 0.366 & 0.244 & 0.253 & 0.319 & 0.350 \\
24 & 0.371 & 0.404 & 0.436 & 0.403 & 0.276 & 0.258 & 0.330 & 0.351 \\
25 & 0.349 & 0.423 & 0.443 & 0.412 & 0.281 & 0.261 & 0.363 & 0.369 \\
26 & 0.325 & 0.400 & 0.409 & 0.406 & 0.272 & 0.259 & 0.338 & 0.355 \\
\hline
$\langle P_{C} \rangle$ & 0.328 & 0.379 & 0.395 & 0.378 & 0.257 & 0.250 & 
0.337 & 0.331 \\
$\sigma(P_{C})$ & 0.041 & 0.046 & 0.046 & 0.036 & 0.021 & 0.017 & 0.036 & 
0.034 \\
\hline
\end{tabular}
\end{table}

\clearpage

\begin{table}[htbp]
\centering
\caption{Probabilities $P(XZA)$ for the eight amino-acids related to 
quartets (release 131.0 of GenBank database). First column refers to the 
biological species of table \ref{tabledata1}.}
\label{tablePA}
\footnotesize \medskip
\begin{tabular}{|c||c|c|c|c|c|c|c|c|}
\hline
& $P_{A}(P)$ & $P_{A}(T)$ & $P_{A}(A)$ & $P_{A}(S)$ & $P_{A}(V)$ & 
$P_{A}(L)$ & $P_{A}(R)$ & $P_{A}(G)$ \\
\hline
1 & 0.273 & 0.282 & 0.228 & 0.246 & 0.117 & 0.097 & 0.189 & 0.248 \\
2 & 0.284 & 0.291 & 0.229 & 0.232 & 0.116 & 0.098 & 0.213 & 0.256 \\
3 & 0.275 & 0.278 & 0.221 & 0.225 & 0.110 & 0.092 & 0.205 & 0.251 \\
4 & 0.361 & 0.346 & 0.317 & 0.257 & 0.174 & 0.134 & 0.241 & 0.345 \\
5 & 0.269 & 0.289 & 0.248 & 0.237 & 0.119 & 0.084 & 0.163 & 0.262 \\
6 & 0.282 & 0.294 & 0.246 & 0.248 & 0.103 & 0.083 & 0.225 & 0.341 \\
7 & 0.261 & 0.273 & 0.209 & 0.237 & 0.111 & 0.111 & 0.198 & 0.250 \\
8 & 0.264 & 0.257 & 0.198 & 0.230 & 0.108 & 0.131 & 0.188 & 0.250 \\
9 & 0.307 & 0.315 & 0.261 & 0.271 & 0.141 & 0.107 & 0.206 & 0.281 \\
10 & 0.224 & 0.224 & 0.180 & 0.182 & 0.078 & 0.060 & 0.155 & 0.224 \\
11 & 0.256 & 0.265 & 0.203 & 0.224 & 0.115 & 0.091 & 0.183 & 0.256 \\
12 & 0.239 & 0.211 & 0.180 & 0.180 & 0.073 & 0.059 & 0.183 & 0.268 \\
13 & 0.230 & 0.250 & 0.199 & 0.196 & 0.102 & 0.098 & 0.148 & 0.277 \\
14 & 0.279 & 0.286 & 0.221 & 0.233 & 0.105 & 0.087 & 0.178 & 0.254 \\
15 & 0.241 & 0.233 & 0.184 & 0.206 & 0.089 & 0.083 & 0.179 & 0.248 \\
16 & 0.262 & 0.262 & 0.224 & 0.230 & 0.079 & 0.077 & 0.178 & 0.324 \\
17 & 0.283 & 0.282 & 0.234 & 0.222 & 0.121 & 0.097 & 0.211 & 0.247 \\
18 & 0.268 & 0.273 & 0.216 & 0.219 & 0.103 & 0.087 & 0.208 & 0.251 \\
19 & 0.289 & 0.271 & 0.217 & 0.231 & 0.087 & 0.077 & 0.167 & 0.284 \\
20 & 0.288 & 0.285 & 0.221 & 0.226 & 0.106 & 0.084 & 0.205 & 0.254 \\
21 & 0.204 & 0.216 & 0.160 & 0.179 & 0.060 & 0.086 & 0.157 & 0.217 \\
22 & 0.309 & 0.297 & 0.236 & 0.270 & 0.097 & 0.096 & 0.221 & 0.337 \\
23 & 0.284 & 0.277 & 0.216 & 0.215 & 0.114 & 0.095 & 0.208 & 0.246 \\
24 & 0.242 & 0.241 & 0.199 & 0.214 & 0.109 & 0.110 & 0.198 & 0.255 \\
25 & 0.239 & 0.238 & 0.189 & 0.223 & 0.102 & 0.098 & 0.178 & 0.241 \\
26 & 0.262 & 0.246 & 0.194 & 0.188 & 0.096 & 0.086 & 0.205 & 0.238 \\
\hline
$\langle P_{A} \rangle$ & 0.268 & 0.268 & 0.217 & 0.224 & 0.105 & 0.093 & 
0.192 & 0.266 \\
$\sigma(P_{A})$ & 0.031 & 0.031 & 0.031 & 0.025 & 0.022 & 0.017 & 0.023 & 
0.034 \\
\hline
\end{tabular}
\end{table}

\clearpage

\begin{table}[htbp]
\centering
\caption{Probabilities $P(XZU)$ for the eight amino-acids related to 
quartets (release 131.0 of GenBank database). First column refers to the 
biological species of table \ref{tabledata1}.}
\label{tablePU}
\footnotesize \medskip
\begin{tabular}{|c||c|c|c|c|c|c|c|c|}
\hline
& $P_{U}(P)$ & $P_{U}(T)$ & $P_{U}(A)$ & $P_{U}(S)$ & $P_{U}(V)$ & 
$P_{U}(L)$ & $P_{U}(R)$ & $P_{U}(G)$ \\
\hline
1 & 0.283 & 0.242 & 0.263 & 0.302 & 0.179 & 0.162 & 0.140 & 0.163 \\
2 & 0.304 & 0.247 & 0.291 & 0.318 & 0.168 & 0.160 & 0.149 & 0.175 \\
3 & 0.295 & 0.233 & 0.280 & 0.303 & 0.156 & 0.149 & 0.152 & 0.169 \\
4 & 0.320 & 0.300 & 0.333 & 0.367 & 0.269 & 0.253 & 0.259 & 0.215 \\
5 & 0.257 & 0.234 & 0.281 & 0.289 & 0.196 & 0.155 & 0.167 & 0.168 \\
6 & 0.300 & 0.251 & 0.308 & 0.321 & 0.214 & 0.170 & 0.221 & 0.217 \\
7 & 0.266 & 0.215 & 0.256 & 0.291 & 0.164 & 0.144 & 0.137 & 0.160 \\
8 & 0.244 & 0.197 & 0.240 & 0.255 & 0.139 & 0.121 & 0.119 & 0.139 \\
9 & 0.303 & 0.273 & 0.288 & 0.326 & 0.209 & 0.199 & 0.167 & 0.177 \\
10 & 0.238 & 0.193 & 0.220 & 0.245 & 0.138 & 0.118 & 0.113 & 0.132 \\
11 & 0.280 & 0.231 & 0.260 & 0.307 & 0.157 & 0.150 & 0.127 & 0.179 \\
12 & 0.241 & 0.193 & 0.259 & 0.253 & 0.168 & 0.114 & 0.186 & 0.186 \\
13 & 0.282 & 0.220 & 0.291 & 0.320 & 0.161 & 0.116 & 0.267 & 0.220 \\
14 & 0.277 & 0.223 & 0.267 & 0.285 & 0.169 & 0.142 & 0.135 & 0.178 \\
15 & 0.263 & 0.214 & 0.246 & 0.289 & 0.146 & 0.132 & 0.151 & 0.158 \\
16 & 0.294 & 0.230 & 0.303 & 0.310 & 0.243 & 0.164 & 0.196 & 0.197 \\
17 & 0.318 & 0.256 & 0.322 & 0.344 & 0.173 & 0.169 & 0.172 & 0.202 \\
18 & 0.302 & 0.227 & 0.282 & 0.313 & 0.154 & 0.143 & 0.156 & 0.166 \\
19 & 0.281 & 0.238 & 0.273 & 0.306 & 0.173 & 0.162 & 0.167 & 0.162 \\
20 & 0.292 & 0.237 & 0.282 & 0.308 & 0.156 & 0.154 & 0.170 & 0.177 \\
21 & 0.218 & 0.166 & 0.257 & 0.321 & 0.116 & 0.105 & 0.105 & 0.109 \\
22 & 0.305 & 0.281 & 0.344 & 0.334 & 0.226 & 0.171 & 0.286 & 0.252 \\
23 & 0.296 & 0.251 & 0.272 & 0.311 & 0.154 & 0.150 & 0.151 & 0.169 \\
24 & 0.265 & 0.212 & 0.243 & 0.285 & 0.149 & 0.133 & 0.127 & 0.150 \\
25 & 0.305 & 0.220 & 0.255 & 0.276 & 0.147 & 0.126 & 0.163 & 0.168 \\
26 & 0.302 & 0.237 & 0.297 & 0.317 & 0.155 & 0.140 & 0.169 & 0.180 \\
\hline
$\langle P_{U} \rangle$ & 0.282 & 0.232 & 0.277 & 0.304 & 0.172 & 0.150 & 
0.167 & 0.176 \\
$\sigma(P_{U})$ & 0.026 & 0.029 & 0.029 & 0.028 & 0.035 & 0.030 & 0.046 & 
0.030 \\
\hline
\end{tabular}
\end{table}

\clearpage

\begin{table}[htbp]
\centering
\caption{Probabilities $P(XZG)$ for the eight amino-acids related to 
quartets (release 131.0 of GenBank database). First column refers to the 
biological species of table \ref{tabledata1}.}
\label{tablePG}
\footnotesize \medskip
\begin{tabular}{|c||c|c|c|c|c|c|c|c|}
\hline
& $P_{G}(P)$ & $P_{G}(T)$ & $P_{G}(A)$ & $P_{G}(S)$ & $P_{G}(V)$ & 
$P_{G}(L)$ & $P_{G}(R)$ & $P_{G}(G)$ \\
\hline
1 & 0.114 & 0.114 & 0.106 & 0.091 & 0.465 & 0.495 & 0.349 & 0.248 \\
2 & 0.105 & 0.108 & 0.097 & 0.087 & 0.466 & 0.494 & 0.331 & 0.235 \\
3 & 0.111 & 0.117 & 0.099 & 0.092 & 0.473 & 0.504 & 0.331 & 0.240 \\
4 & 0.084 & 0.081 & 0.072 & 0.075 & 0.358 & 0.416 & 0.238 & 0.204 \\
5 & 0.143 & 0.144 & 0.124 & 0.115 & 0.455 & 0.527 & 0.306 & 0.249 \\
6 & 0.164 & 0.143 & 0.134 & 0.118 & 0.445 & 0.509 & 0.226 & 0.157 \\
7 & 0.124 & 0.129 & 0.108 & 0.095 & 0.468 & 0.491 & 0.336 & 0.241 \\
8 & 0.136 & 0.131 & 0.114 & 0.096 & 0.487 & 0.494 & 0.340 & 0.245 \\
9 & 0.099 & 0.097 & 0.089 & 0.075 & 0.420 & 0.452 & 0.338 & 0.243 \\
10 & 0.154 & 0.168 & 0.125 & 0.126 & 0.507 & 0.553 & 0.339 & 0.252 \\
11 & 0.113 & 0.125 & 0.114 & 0.093 & 0.464 & 0.500 & 0.350 & 0.231 \\
12 & 0.182 & 0.195 & 0.152 & 0.150 & 0.454 & 0.561 & 0.268 & 0.214 \\
13 & 0.104 & 0.100 & 0.093 & 0.080 & 0.455 & 0.536 & 0.204 & 0.204 \\
14 & 0.109 & 0.112 & 0.099 & 0.090 & 0.481 & 0.520 & 0.323 & 0.223 \\
15 & 0.127 & 0.123 & 0.118 & 0.103 & 0.499 & 0.508 & 0.299 & 0.233 \\
16 & 0.151 & 0.136 & 0.122 & 0.088 & 0.409 & 0.525 & 0.259 & 0.179 \\
17 & 0.079 & 0.083 & 0.067 & 0.073 & 0.463 & 0.499 & 0.320 & 0.209 \\
18 & 0.115 & 0.118 & 0.101 & 0.095 & 0.469 & 0.514 & 0.327 & 0.240 \\
19 & 0.112 & 0.108 & 0.107 & 0.073 & 0.478 & 0.492 & 0.331 & 0.244 \\
20 & 0.088 & 0.099 & 0.093 & 0.090 & 0.481 & 0.525 & 0.317 & 0.246 \\
21 & 0.163 & 0.122 & 0.194 & 0.081 & 0.582 & 0.535 & 0.320 & 0.307 \\
22 & 0.127 & 0.111 & 0.106 & 0.087 & 0.431 & 0.500 & 0.195 & 0.138 \\
23 & 0.124 & 0.114 & 0.117 & 0.108 & 0.487 & 0.502 & 0.323 & 0.235 \\
24 & 0.122 & 0.143 & 0.122 & 0.099 & 0.466 & 0.499 & 0.346 & 0.244 \\
25 & 0.108 & 0.119 & 0.114 & 0.089 & 0.470 & 0.516 & 0.296 & 0.222 \\
26 & 0.111 & 0.117 & 0.100 & 0.089 & 0.476 & 0.515 & 0.289 & 0.227 \\
\hline
$\langle P_{G} \rangle$ & 0.122 & 0.121 & 0.111 & 0.095 & 0.466 & 0.507 & 
0.304 & 0.227 \\
$\sigma(P_{G})$ & 0.025 & 0.024 & 0.025 & 0.017 & 0.038 & 0.029 & 0.045 & 
0.033 \\
\hline
\end{tabular}
\end{table}

\clearpage

\begin{table}[htbp]
\centering
\caption{Probabilities $P(XZC)+P(XZA)$ for the eight amino-acids related to 
quartets (release 131.0 of GenBank database). First column refers to the 
biological species of table \ref{tabledata1}.}
\label{tablePCA}
\footnotesize \medskip
\begin{tabular}{|c||c|c|c|c|c|c|c|c|}
\hline
& $P_{C+A}(P)$ & $P_{C+A}(T)$ & $P_{C+A}(A)$ & $P_{C+A}(S)$ & $P_{C+A}(V)$ 
& $P_{C+A}(L)$ & $P_{C+A}(R)$ & $P_{C+A}(G)$ \\
\hline
1 & 0.603 & 0.644 & 0.631 & 0.607 & 0.356 & 0.343 & 0.512 & 0.589 \\
2 & 0.591 & 0.645 & 0.612 & 0.595 & 0.366 & 0.345 & 0.519 & 0.590 \\
3 & 0.593 & 0.650 & 0.621 & 0.605 & 0.371 & 0.347 & 0.518 & 0.591 \\
4 & 0.596 & 0.619 & 0.595 & 0.559 & 0.373 & 0.331 & 0.503 & 0.582 \\
5 & 0.600 & 0.622 & 0.595 & 0.596 & 0.350 & 0.318 & 0.527 & 0.583 \\
6 & 0.536 & 0.606 & 0.557 & 0.561 & 0.341 & 0.321 & 0.552 & 0.626 \\
7 & 0.610 & 0.656 & 0.636 & 0.614 & 0.368 & 0.365 & 0.527 & 0.599 \\
8 & 0.620 & 0.671 & 0.646 & 0.649 & 0.374 & 0.386 & 0.541 & 0.615 \\
9 & 0.599 & 0.630 & 0.623 & 0.600 & 0.370 & 0.349 & 0.496 & 0.581 \\
10 & 0.608 & 0.640 & 0.654 & 0.629 & 0.355 & 0.329 & 0.548 & 0.616 \\
11 & 0.607 & 0.644 & 0.627 & 0.600 & 0.379 & 0.350 & 0.523 & 0.590 \\
12 & 0.577 & 0.613 & 0.589 & 0.598 & 0.378 & 0.325 & 0.546 & 0.600 \\
13 & 0.614 & 0.680 & 0.616 & 0.600 & 0.385 & 0.348 & 0.529 & 0.575 \\
14 & 0.614 & 0.666 & 0.633 & 0.625 & 0.350 & 0.338 & 0.542 & 0.599 \\
15 & 0.610 & 0.663 & 0.636 & 0.607 & 0.355 & 0.360 & 0.550 & 0.609 \\
16 & 0.555 & 0.634 & 0.575 & 0.601 & 0.348 & 0.311 & 0.545 & 0.623 \\
17 & 0.603 & 0.661 & 0.611 & 0.583 & 0.364 & 0.331 & 0.507 & 0.588 \\
18 & 0.584 & 0.655 & 0.618 & 0.592 & 0.377 & 0.343 & 0.517 & 0.594 \\
19 & 0.607 & 0.654 & 0.620 & 0.621 & 0.349 & 0.345 & 0.501 & 0.594 \\
20 & 0.619 & 0.663 & 0.625 & 0.602 & 0.363 & 0.321 & 0.513 & 0.577 \\
21 & 0.619 & 0.712 & 0.548 & 0.598 & 0.302 & 0.359 & 0.575 & 0.584 \\
22 & 0.568 & 0.609 & 0.550 & 0.579 & 0.344 & 0.329 & 0.519 & 0.609 \\
23 & 0.581 & 0.635 & 0.611 & 0.581 & 0.358 & 0.348 & 0.526 & 0.596 \\
24 & 0.614 & 0.645 & 0.635 & 0.617 & 0.385 & 0.368 & 0.527 & 0.606 \\
25 & 0.588 & 0.661 & 0.631 & 0.635 & 0.382 & 0.359 & 0.541 & 0.610 \\
26 & 0.587 & 0.646 & 0.603 & 0.594 & 0.368 & 0.345 & 0.542 & 0.593 \\
\hline
$\langle P_{C+A} \rangle$ & 0.596 & 0.647 & 0.611 & 0.602 & 0.362 & 0.343 & 
0.529 & 0.597 \\
$\sigma(P_{C+A})$ & 0.021 & 0.023 & 0.028 & 0.021 & 0.018 & 0.017 & 0.019 & 
0.014 \\
\hline
\end{tabular}
\end{table}

\clearpage

\begin{table}[htbp]
\centering
\caption{Shannon entropy for the eight amino-acids related to quartets 
(release 131.0 of GenBank database). First column refers to the biological 
species of table \ref{tabledata1}.}
\label{shannon}
\footnotesize \medskip
\begin{tabular}{|c||c|c|c|c|c|c|c|c|}
\hline
& $S(P)$ & $S(T)$ & $S(A)$ & $S(S)$ & $S(V)$ & $S(L)$ & $S(R)$ & $S(G)$ \\
\hline
& Pro & Thr & Ala & Ser & Val & Leu & Arg & Gly \\
1 & 1.911 & 1.898 & 1.865 & 1.865 & 1.815 & 1.751 & 1.908 & 1.953 \\
2 & 1.903 & 1.893 & 1.862 & 1.852 & 1.806 & 1.752 & 1.936 & 1.963 \\
3 & 1.910 & 1.896 & 1.854 & 1.853 & 1.785 & 1.726 & 1.933 & 1.957 \\
4 & 1.847 & 1.857 & 1.841 & 1.835 & 1.943 & 1.879 & 1.999 & 1.966 \\
5 & 1.943 & 1.939 & 1.916 & 1.900 & 1.831 & 1.694 & 1.911 & 1.964 \\
6 & 1.966 & 1.945 & 1.934 & 1.913 & 1.827 & 1.721 & 1.978 & 1.943 \\
7 & 1.917 & 1.900 & 1.846 & 1.864 & 1.796 & 1.760 & 1.912 & 1.948 \\
8 & 1.926 & 1.877 & 1.832 & 1.841 & 1.757 & 1.758 & 1.878 & 1.924 \\
9 & 1.893 & 1.887 & 1.864 & 1.845 & 1.884 & 1.821 & 1.947 & 1.973 \\
10 & 1.922 & 1.900 & 1.812 & 1.840 & 1.691 & 1.590 & 1.830 & 1.900 \\
11 & 1.903 & 1.902 & 1.853 & 1.856 & 1.801 & 1.730 & 1.885 & 1.964 \\
12 & 1.964 & 1.920 & 1.892 & 1.883 & 1.747 & 1.575 & 1.939 & 1.965 \\
13 & 1.873 & 1.837 & 1.827 & 1.807 & 1.792 & 1.671 & 1.915 & 1.983 \\
14 & 1.904 & 1.883 & 1.848 & 1.847 & 1.779 & 1.697 & 1.891 & 1.958 \\
15 & 1.911 & 1.860 & 1.828 & 1.854 & 1.723 & 1.693 & 1.908 & 1.940 \\
16 & 1.956 & 1.916 & 1.906 & 1.852 & 1.823 & 1.691 & 1.940 & 1.954 \\
17 & 1.856 & 1.846 & 1.809 & 1.819 & 1.816 & 1.751 & 1.957 & 1.966 \\
18 & 1.914 & 1.891 & 1.855 & 1.858 & 1.777 & 1.705 & 1.940 & 1.954 \\
19 & 1.912 & 1.880 & 1.864 & 1.817 & 1.761 & 1.723 & 1.919 & 1.962 \\
20 & 1.873 & 1.870 & 1.844 & 1.852 & 1.774 & 1.696 & 1.952 & 1.969 \\
21 & 1.900 & 1.780 & 1.916 & 1.790 & 1.553 & 1.641 & 1.813 & 1.880 \\
22 & 1.929 & 1.910 & 1.889 & 1.868 & 1.833 & 1.750 & 1.978 & 1.936 \\
23 & 1.928 & 1.901 & 1.880 & 1.878 & 1.775 & 1.733 & 1.935 & 1.953 \\
24 & 1.903 & 1.899 & 1.852 & 1.850 & 1.784 & 1.743 & 1.897 & 1.940 \\
25 & 1.892 & 1.864 & 1.833 & 1.833 & 1.769 & 1.702 & 1.921 & 1.940 \\
26 & 1.908 & 1.881 & 1.838 & 1.817 & 1.762 & 1.699 & 1.948 & 1.954 \\
\hline
$\langle S \rangle$ & 1.910 & 1.886 & 1.860 & 1.850 & 1.785 & 1.717 & 1.922 
& 1.950 \\
$\sigma(S)$ & 0.029 & 0.033 & 0.032 & 0.027 & 0.068 & 0.061 & 0.041 & 0.022 
\\
\hline
\end{tabular}
\end{table}

\end{document}